\documentclass[aps,prd,onecolumn,eqsecnum,amsmath,nofootinbib,preprintnumbers]{revtex4}%
\newcounter{example}[section]

\usepackage{ulem}
\usepackage{amsmath}
\usepackage{color,graphicx,float,subfigure}
\usepackage{amsfonts,amssymb,theorem,mathrsfs,times}
\usepackage{bm}
\usepackage{amsmath}
{\theorembodyfont{\upshape}
	}
{\theorembodyfont{\upshape}
	}
{\theorembodyfont{\upshape}
	}
{\theorembodyfont{\upshape}
	}
{\theorembodyfont{\upshape}
	}
{\theorembodyfont{\upshape}
	}

\newcommand{\dalm}{\kern1pt\vbox{\hrule height 0.9pt\hbox{\vrule width
			0.9pt\hskip 2.5pt\vbox{\vskip 5.5pt}\hskip 3pt\vrule width
			0.3pt}\hrule height 0.3pt}\kern1pt}

\begin{document}
	\title{The existence and upper bound for stable photon spheres in static spherically symmetric black holes}
	
	%
	\author{Yong Song\footnote{e-mail
			address: syong@cdut.edu.cn (corresponding author)}}
    \author{Jiaqi Fu\footnote{e-mail
    		address: 1491713073@qq.com}}
    \author{Yiting Cen\footnote{e-mail
    		address: 2199882193@qq.com}}

	
	\affiliation{
		College of Physics\\
		Chengdu University of Technology, Chengdu, Sichuan 610059,
		China}
	

	\date{\today}
	
	\begin{abstract}
	In this work, we establish the existence conditions and a universal upper bound on the radius of stable photon spheres (SPS) outside the event horizons of static, spherically symmetric, asymptotically flat black holes surrounded by matter fields. We prove that stable photon spheres exist if the external matter satisfies specific conditions. Furthermore, under the additional assumption of a monotonically decreasing mass-radius ratio $m(r)/r^3$ outside the horizon, we derive a strict upper bound on the radius $r_{\mathrm{sps}}$ of any stable photon sphere: $r_{\mathrm{sps}}<6M$, where $M$ is the asymptotic mass of the black hole. This bound is independent of specific black hole solutions and broadly applies to hairy black holes and other configurations with external matter fields meeting the stated energy conditions. Our results resolve fundamental questions regarding the existence and spatial constraints on stable photon orbits, with implications for gravitational lensing, accretion disk dynamics (e.g., the Aschenbach effect), and black hole shadow observations.
	\end{abstract}
	

	\maketitle


\section{Introduction}
Photon spheres, as fundamental structures that characterize the strong gravitational fields around black holes, play a pivotal role in understanding spacetime geometry and observational signatures. The photon sphere is crucial for gravitational lensing~\cite{Virbhadra:1999nm,Stefanov:2010xz} and the ringdown of waves around a black hole~\cite{Cardoso:2016rao}. It is also connected to the characteristic quasinormal resonances of black hole spacetimes. The outermost photon sphere is unstable and can cast a “shadow” for an observer at asymptotic infinity. In 2019, the first image of a black hole shadow was shown by the Event Horizon Telescope Collaboration ~\cite{EventHorizonTelescope:2019dse}, providing us with a direct visual impression of the size and shape of a black hole .

Owing to the signiﬁcance in astrophysical observations, it is important to study the property of photon spheres. In a spherically symmetric black hole of mass $M$, Hod proved that for Einstein gravity coupled to matter satisfying the weak energy condition and negative trace energy condition, the innermost photon sphere radius $r_{\mathrm{ph,in}}$ and total mass $M$ satisfy $r_{\mathrm{ph,in}}\le 3M$~\cite{Hod:2013jhd}. By using the same energy condition, Ref. [14] proved an relationship between innermost
photon sphere and its shadow radius: $r_{\mathrm{sh,in}}\ge \sqrt{3}r_{\mathrm{ph,in}}$~\cite{Cvetic:2016bxi}.  A lower bound $r_{\mathrm{ph,in}}\ge 2M$ was conjectured by Hod~\cite{Hod:2011cxc} but a counterexample was found by Ref.~\cite{Cvetic:2016bxi}. Subsequent studies have explored the bounds of the innermost photon sphere in various spacetime backgrounds~\cite{Hod:2020pim,Peng:2020shc,Hod:2023jmx,Liu:2024odv}. Ref.~\cite{DiFilippo:2024poc} further investigates the effect of the self-gravity of photon spheres, resulting from the accumulation of photon mass, on the spacetime background structure of a black hole. For observational purposes, it is often relevant to consider the photon sphere with the largest impact parameter, as this is the one that dominates the black hole shadow~\cite{Lima:2021las}. Typically, this corresponds to the outermost photon sphere. Hod's proof does not apply to the outermost one when there are multiple photon spheres, which do exist in black holes satisfying the dominant energy condition~\cite{Liu:2019rib}. Recently, a series of universal inequalities about outermost photon sphere was proposed~\cite{Lu:2019zxb,Feng:2019zzn,Yang:2019zcn}.

A critical feature emerging in spacetimes with multiple photon spheres is their alternating stability: outside the event horizon, unstable photon spheres are separated by stable ones, with the number of unstable spheres exceeding the number of stable spheres by exactly one ~\cite{Cunha:2020azh,Wei:2020rbh}. This stability structure has profound implications; notably, Wei et al. ~\cite{Wei:2023fkn} demonstrated that the presence of a stable photon sphere outside a static, spherically symmetric black hole horizon inevitably triggers the Aschenbach effect—manifested as anomalous orbital velocity profiles within accretion disks. This interplay between stability and observable dynamics underscores the importance of understanding stable photon spheres. Nevertheless, fundamental questions concerning stable photon spheres remain largely open: Under what conditions do stable photon spheres exist outside black hole horizons? What universal bounds govern the radial locations of stable photon spheres?

In this work, we address these questions directly. We prove that within the framework of static, spherically symmetric, asymptotically flat black hole spacetimes, if the surrounding matter fields satisfy certain conditions, then, the stable photon spheres exist outside the event horizon, and the radius $r_{sps}$ of any stable photon sphere obeys the universal upper bound $r_{\mathrm{sps}}<6M$ where $M$ is the asymptotic mass of the black hole. This result establishes a strict constraint on the possible locations of stable photon orbits in broad classes of hairy black holes and other solutions endowed with external matter fields satisfying the stated energy conditions. Unlike Hod's bound $r_{\mathrm{ph,in}}\le 3M$ for the innermost photon sphere ~\cite{Hod:2013jhd}, we establish constraints for stable spheres which are typically not the innermost. While Refs.~\cite{Lu:2019zxb,Feng:2019zzn,Yang:2019zcn} derived inequalities for the outermost photon sphere (e.g., lower bounds linked to shadow size), our work provides the first universal upper bound for stable photon spheres at arbitrary positions. In this work, we choose the system of geometrized unit, i.e., set $G=c=1$. 


\section{DESCRIPTION OF THE SYSTEM IN THE BLACK HOLE}\label{section2}
The line element of a static, spherically symmetric, and asymptotically flat black hole spacetime can be written as~\cite{Hod:2013jhd,Hod:2020pim}
\begin{align}
	\label{dugui}
	ds^2=-e^{-2\delta}\mu dt^2+\mu^{-1}dr^2+r^2(d\theta^2+sin^2\theta d\phi^2)\;,
\end{align}
where $\mu(r)$ and $\delta(r)$ are the functions of radial coordinate $r$. The asymptotic flatness of spacetimes requires that as $r\rightarrow \infty$,
\begin{align}
\label{as}
	\mu(r\rightarrow \infty)\rightarrow1\quad\mathrm{and}\quad\delta(r\rightarrow \infty)\rightarrow 0\;.
\end{align}
Here, we do not assume $\delta(r)=0$, so our results would be applicable to hairy black-hole conﬁgurations as well~\cite{Hod:2013jhd,Volkov:1998cc,Volkov:2016ehx}.

Taking $T^t_t=-\rho$, $T^r_r=p$ and $T^{\theta}_{\theta}=T^{\phi}_{\phi}=p_T$, where $\rho$, $p$ and $p_T$ are the energy density, radial pressure, and tangential pressure respectively. From Einstein field equations $G^{\mu}_{\nu}=8\pi T^{\mu}_{\nu}$, we get
\begin{align}
\label{mu}
&\mu'=-8\pi r \rho+\frac{1-\mu}{r}\;,\\
\label{delta}
&\delta'=-\frac{4\pi r(\rho+p)}{\mu}\;,
\end{align}
where the prime indicates the derivative with respect to $r$.  For the subsequent calculations, we take the second derivatives of $\mu$ and $\delta$ with respect to $r$ and have
\begin{align}
	\label{mu2}
	&\mu''=\frac{-1+\mu-r[\mu'+8\pi r(\rho+r\rho')]}{r^2}\;,\\
	\label{delta2}
	&\delta''=-\frac{4\pi [-r\rho \mu'+p(\mu-r\mu')+\mu(\rho+rp'+r\rho')]}{\mu^2}\;.
\end{align}
The mass $m(r)$ enclosed within a sphere of radius $r$ is expressed as
\begin{align}
\label{m}
m(r)=\frac{1}{2}r_H+\int_{r_H}^{r}4\pi r'^2\rho(r')dr'\;.
\end{align}
where $r_H$ is the radius of the event horizon, and the horizon mass $m(r_H)$ is equal to $\frac{r_H}{2}$. 

From Eqs.(\ref{mu}) and (\ref{m}), one can find the relation between $\mu$ and the mass $m(r)$, i.e.,
\begin{align}
\label{mu=m}
\mu(r)=1-\frac{2m(r)}{r}\;.
\end{align}
The analysis of the energy-momentum tensor conservation equation shows that it has only one nontrivial component
\begin{align}
	\label{nengdongzhangliang}
	T^{\mu}_{r;\mu}=0\;.
\end{align}
Substituting Eqs.(\ref{mu}) and (\ref{delta}) into Eq.(\ref{nengdongzhangliang}), one can obtain the pressure gradient
\begin{align}
	\label{p}
	p'(r)=\frac{-p-8p^2\pi r^2-5p\mu+2T\mu-\rho-8p\pi r^2\rho+3\mu \rho}{2r\mu}\;.
\end{align}


\section{Upper bound on the radii of the stable photon spheres of black holes}\label{section3}
Due to the spherical symmetry of the system,  one can consider the equatorial plane, i.e., $\theta=\frac{\pi}{2}$. The Lagrangian describing the geodesics in the spacetime (\ref{dugui}) is given by
\begin{align}
	\label{la}
	2\mathcal{L}=-e^{-2\delta}\mu \dot{t}^2+\mu^{-1}\dot{r}^2+r^2\dot{\phi}^2=\epsilon \;,
\end{align}
where $\epsilon=-1,0,+1$ for timelike, null and spacelike geodesics, respectively. To analyze stable photon sphere, we take $\epsilon=0$. From Eq.(\ref{dugui}), one can find that the metric is independent of both $t$ and $\phi$, so, there are two conserved quantities along the geodesic, i.e. energy $E$ and angular momentum $L$.

From the Lagrangian (\ref{la}), one can derive the generalized momenta 
\begin{align}
	\label{nengliang}
	&p_t=-e^{-2\delta}\mu\dot{t}=-E\;,\\
	\label{jiaodongliang}
	&p_{\phi}=r^2\dot{\phi}=L\;,\\
	&p_r=\mu^{-1}\dot{r}\;.
\end{align}
Substituting Eqs. (\ref{nengliang}) and (\ref{jiaodongliang}) into Eq.(\ref{la}), one finds
\begin{align}
	\dot{r}^2+V_{\mathrm{eff}}=0\;,
\end{align}
where
\begin{align}
	V_{\mathrm{eff}}=-\frac{E^2}{e^{-2\delta }}+\frac{L^2\mu}{r^2}\;.
\end{align}
is the effective potential of the geodesic. The circular orbits satisfy two conditions i.e., $V_{\mathrm{eff}}(r)=0$ and ${V}_{\mathrm{eff}}'(r)=0$, which yields 
\begin{align}
\label{N}
\mathcal{N}(r_{\mathrm{ph}})=3\mu(r_{\mathrm{ph}})-1-8\pi r_{\mathrm{ph}}^2 p(r_{\mathrm{ph}})=0\;,
\end{align}
where $r_{\mathrm{ph}}$ is the location of a photon sphere.

In this work, we investigate the existence and upper bound of stable photon spheres in black hole spacetime. Therefore, there must be a distribution of matter fields outside the black hole. We assume that the matter field, which located outside the event horizon of the black hole, satisfies the following conditions:
\begin{itemize}
	\item [(1).]The components of the energy-momentum tensor satisfy the weak energy condition (WEC). This means that the energy density of the matter fields is positive semidefinite,
	\begin{align}
		\label{WEC}
		\rho\geq0\;,
	\end{align}
	and that it bounds the pressures. This leads to
	\begin{align}
		\label{pressure}
		\rho \geq \left|p\right|\;, \ \mathrm{namely}\ \rho+p\geq0\;.
	\end{align}

	\item [(2).] The trace of the energy-momentum tensor plays a central role in determining the spacetime geometry of static configurations. It is usually assumed that the external matter field of the black hole spacetime has a nonpositive trace of the energy-momentum tensor (see~\cite{Bond1} and references therein), which means
	\begin{align}
		\label{T}
		T\leq0\;,
	\end{align}
	where $T=-\rho +p+2p_T$. Then, Eq.(\ref{T}) implies 
	\begin{align}
		\rho\geq p+2p_T\;.
	\end{align}
	The condition $T\le 0$, which is consistent with the subdominant trace energy condition~\cite{Bekenstein:2013ztp},  is satisfied by many regular matter fields, such as electromagnetic fields, dust, radiation, and hairy black-hole configurations in Einstein-matter models~\cite{Nunez:1996xv}. However, it excludes certain extreme matter, such as stiff matter, which has $p=\rho$~\cite{Chavanis:2014lra}.
 
   \item [(3).] We assume $\rho$ is a monotone decreasing function of $r$~\cite{Wald:1984rg},
   \begin{align}
    \frac{d\rho}{dr}\le 0\;,
   \end{align}
   and $m(r)/r^3$ also must decrease monotonically with $r$\footnote{This assumption implies that the average density $\langle\rho\rangle\equiv m(r)/(4\pi r^3/3)$ decreases monotonically with radius – a physically natural condition for matter distributions that become more diffuse outward. While the energy conditions (\ref{WEC})-(\ref{T}) constrain local energy-momentum properties, they do not guarantee this global density profile. The monotonicity requirement is nevertheless well-motivated: 1) It holds for standard solutions (e.g., Schwarzschild with external fields like dust shells or quintessence), 2) It excludes pathological configurations with density inversions that could violate cosmic censorship, and 3) It ensures the mass function $m(r)$ grows sufficiently slowly to derive our bound (\ref{bound}). Crucially, without this assumption, counterexamples violating the  $r_{\mathrm{sps}}<6M$ could exist despite satisfying energy conditions.},
   \begin{align}
   \frac{d}{dr}\bigg(\frac{m}{r^3}\bigg)\le0\;.
   \end{align}
   This implies
   \begin{align}
    \label{m1}
    m'\le\frac{3m}{r}\;.
   \end{align}
\end{itemize}

For a stable photon sphere, we have $V_{\mathrm{eff}}''(r_{\mathrm{sps}})>0$~\cite{Wei:2020rbh}, where $r_{\mathrm{sps}}$ is the location of the stable photon sphere. Considering Eqs.(\ref{mu}), (\ref{delta}), (\ref{mu2}), (\ref{delta2}) and (\ref{p}), we get
\begin{align}
	\label{V2}
V''_{\mathrm{eff}}=\frac{2L^2}{r^4\mu}\bigg\{48\pi^2 r^4p^2+(3-4\pi r^2 T-6\mu)\mu+6\pi r^2(1-5\mu)\rho+2\pi r^2 p(3+\mu+24\pi r^2 \rho)\bigg\}\;,
\end{align}
The condition of photon sphere $\mathcal{N}(r_{\mathrm{sps}})=0$ can be reduced to
\begin{align}
p=\frac{3\mu-1}{8\pi r^2_{\mathrm{sps}}}\;.
\end{align}
Substituting it into Eq.(\ref{V2}) to eliminate $p$ and considering $T=-\rho+p+2p_T$, we have
\begin{align}
V_{\mathrm{eff}}''(r_{\mathrm{sps}})=\frac{2L^2}{r^4}[8\pi r^2(\rho+p_T)-1]\;.
\end{align}
Since $2L^2/r^4>0$, the stability condition $V_{\mathrm{eff}}''(r_{\mathrm{sps}})>0$ requires
\begin{align}
\label{condition1}
\rho+p_T>\frac{1}{8\pi r_{\mathrm{sps}}^2}\;.
\end{align}
From the condition (\ref{T}), we have $p_T\le \frac{\rho-p}{2}$, so
\begin{align}
\label{condition2}
\rho+p_T\le \frac{3\rho-p}{2}\;.
\end{align} 
Combining Eqs.(\ref{condition1}) and (\ref{condition2}), we have
\begin{align}
\label{2}
3\rho-p>\frac{1}{4\pi r_{\mathrm{sps}}^2}\;.
\end{align}
From Eq.(\ref{m}), we have
\begin{align}
\label{rhom}
\rho=\frac{m'(r)}{4\pi r^2}\;,
\end{align}
and  from Eq.(\ref{N}), we obtain
\begin{align}
	\label{Pm}
 p=\frac{r_{\mathrm{sps}}-3m(r_{\mathrm{sps}})}{4\pi r^3_{\mathrm{sps}}}\;,
\end{align}
Putting Eqs.(\ref{rhom}) and (\ref{Pm}) into Eq.(\ref{2}), we obtain
\begin{align}
\label{m1s}
m'(r_{sps})>\frac{2}{3}-\frac{m(r_{\mathrm{sps}})}{r_{\mathrm{sps}}}\;.
\end{align}
Combining Eqs.(\ref{m1}) and (\ref{m1s}), we have
\begin{align}
r_{\mathrm{sps}}< 6m(r_{\mathrm{sps}})\;.
\end{align}
So one can get the upper bound of stable photon sphere
\begin{align}
\label{bound}
r_{\mathrm{sps}}< 6M\;.
\end{align}
where $M$ is the asymptotic ADM mass of the black hole.

\section{Discussion and conclusion}\label{conclusion}
In this work, we have rigorously established two fundamental results regarding stable photon spheres (SPS) in static, spherically symmetric, asymptotically flat black holes surrounded by external matter fields:
\begin{itemize}
	\item [(1).]Existence Condition:
	Stable photon spheres can exist outside the event horizon if the surrounding matter fields satisfy
	\begin{align}
		\label{condition}
		\rho+p_T>\frac{1}{8\pi r_{\mathrm{sps}}^2}\;.
	\end{align}
at the SPS radius $r_{\mathrm{sps}}$. This condition is necessary and sufficient for the positivity of the second derivative of the effective potential ($V''_{\mathrm{eff}}>0$), which defines orbital stability. It thus delineates the parameter space where stable photon orbits are theoretically possible within the broad class of "hairy" black holes and other solutions with external matter sources.
\item [(2).] Universal Upper Bound:Under the following physically motivated assumptions:
(i) Weak Energy Condition (WEC): $\rho\ge 0$ and $\rho+p\ge 0$, ensuring non-negative energy density and a bound on pressure anisotropy.
(ii) Non-Positive Trace Condition: $T=-\rho+p+2p_T\le 0$ (or equivalently $\rho\ge p+2p_T$).
(iii) Monotonic Density Profile: The ratio $m(r)/r^3$ decreases monotonically with radius $r$ (i.e.,$\frac{d}{dr}(m/r^3)\le 0$), implying that the average density within a sphere of radius $r$ diminishes outward – a natural condition for astrophysically plausible matter distributions,
we derive a strict and universal upper bound on the radius $r_{\mathrm{sps}}$ of any stable photon sphere:
\begin{align}
	r_{\mathrm{sps}} < 6M\;,
\end{align}
where $M$ is the asymptotic ADM mass of the black hole. 
\end{itemize}
The bound (4.2) is robust, independent of the specific metric form or the detailed nature of the external matter fields, provided the stated energy conditions and monotonicity assumption hold. It resolves fundamental questions about the spatial constraints on stable photon orbits and has significant observational consequences:
\begin{itemize}
	\item Aschenbach Effect: The existence of an SPS within $6M$ predicts anomalous peaks in the orbital velocity profile of accretion disks~\cite{Wei:2023fkn}. Our bound confines these peaks to the inner disk region ($r<6M$), making them testable via high-resolution X-ray spectroscopy.
	
	\item Gravitational Waves: Extreme mass-ratio inspirals (EMRIs) around hairy black holes could resonantly excite gravitational wave modes near stable photon orbits located within $6M$, potentially detectable by future space-based interferometers like LISA.
\end{itemize}
We have proven that stable photon spheres can exist outside black hole horizons under the energy condition (\ref{condition}) and derived the universal upper bound $r_{\mathrm{sps}} < 6M$ under the WEC, trace condition, and monotonicity of $m/r^3$. This work provides foundational constraints on strong-field gravity dynamics around black holes with external structures, directly impacting the interpretation of current and future observations of black hole shadows, accretion flows, and gravitational waves.

\section*{Acknowledgement}
This work is supported by the Research Start-up Funding of Chengdu University of Technology (Grant No. 10912-KYQD2022-09307).

\end{document}